\documentclass[entropy,article,submit,pdftex,moreauthors]{Definitions/mdpi} 
\renewcommand{\linenumbers}{}

\usepackage{comment}
\firstpage{1} 
\pubvolume{1}
\issuenum{1}
\articlenumber{0}
\pubyear{2022}
\copyrightyear{2022}
\datereceived{} 
\dateaccepted{} 
\datepublished{} 
\hreflink{https://doi.org/} 

\Title{Assigning degrees of stochasticity to blazar light curves in the radio band using complex networks}

\TitleCitation{Title}

\Author{Belén Acosta-Tripailao $^{1,*}$ \orcidA{}, Walter Max-Moerbeck$^{2,*}$ \orcidB{}, Denisse Pastén$^{1,*}$ \orcidC{} and Pablo S. Moya$^{1,*}$\orcidD{}}

\AuthorNames{Belén Acosta-Tripailao, Walter Max-Moerbeck, Denisse Pastén, Pablo S. Moya}

\AuthorCitation{Acosta-Tripailao, B.; Max-Moerbeck, W.; Pastén, D.; Moya, P. S.}

\address{%
$^{1}$ \quad Departamento de Física, Facultad de Ciencias, Universidad de Chile, Las Palmeras 3425, Ñuñoa, Santiago, Chile\\
$^{2}$ \quad Departamento de Astronomía, Facultad de Ciencias Físicas y Matemáticas, Universidad de Chile, Camino El Observatorio 1515, Las Condes, Santiago, Chile}

\corres{Correspondence: bacostaazocar@gmail.com (B.A.); wmax@das.uchile.cl (W.M.); denisse.pasten.g@gmail.com (D.P.); pablo.moya@uchile.cl (P.S.M.)}

\abstract{We focus on characterizing the high-energy emission mechanisms of blazars by analyzing the variability in the radio band of the light curves of more than a thousand sources. We are interested in assigning complexity parameters to these sources, modeling the time series of the light curves with the method of the Horizontal Visibility Graph (HVG), which allows us to obtain properties from degree distributions, such as a characteristic exponent to describe its stochasticity and the Kullback-Leibler Divergence (KLD), presenting a new perspective to the methods commonly used to study Active Galactic Nuclei (AGN). We contrast these parameters with the excess variance, an astronomical measurement of variability in light curves, at the same time we use the spectral classification of the sources. While it is not possible to find significant correlations with the excess variance, the degree distributions extracted from the network are detecting differences related to the spectral classification of blazars. These differences suggest a chaotic behavior in the time series for the BL Lac sources and a correlated stochastic behavior in the time series for the FSRQ sources. Our results show that complex networks may be a valuable alternative tool to study AGNs according to the variability of their energy output.}

\keyword{light curves; blazar; horizontal visibility graph; stochasticity; variability; time series
analysis.}

\begin{document}


\section{Introduction}
\label{sec:intro}
Complex networks are a powerful tool to study physical phenomena in a wide variety of systems and topics from a different perspective than usual approaches \cite{RevModPhys.74.47}. Depending on the characteristics to be explored in each investigation there are different types of graphs and representation structures to consider. Here we are interested in modeling time series with the techniques from the family of visibility algorithms \cite{lacasa2009visibility}. In one of the first approaches to astrophysical systems through the use of Horizontal Visibility Graph (HVG), we have shown that this method is able to detect differences in particle velocity distributions in plasma simulations \cite{acosta2021applying}. Moreover, the HVG has proved to be a robust method to characterize the solar wind plasma and has been used to study turbulent magnetic field \cite{acosta2019reversibility}, velocity fluctuations \cite{suyal2014visibility} and light curves of pulsating variable stars \cite{munoz2021analysis}. Being the closest star to Earth, the Sun and the solar wind are arguably the most studied astrophysical systems, corresponding to a valuable laboratory of natural plasma physics. During the last decades, several space missions have been launched and surveyed the space environment, making many discoveries. In contrast, the study of distant objects, such as blazars, has comparatively fewer high-quality data sets available.

\citet{angel1980optical} indicated that the word blazar was proposed by Edward A. Spiegel in the Pittsburgh Conference on BL Lac Objects in 1978, which is a combination of BL Lacertae object and quasar. Blazars are a particular type of active galactic nuclei (AGN). Emission within an AGN is produced by the accretion of matter from a black hole at its center, where the surrounding material forms an accretion disk that is heated by the dissipation of gravitational energy, generating in some cases the expulsion of matter and energy in relativistic jets. An AGN is powered by the conversion of gravitational potential energy into radiation, although the rotational kinetic energy of the black hole can also serve as an important energy source, moreover, plasma jets are formed when the black hole rotates and the accretion disk is strongly magnetized \cite{blandford2019relativistic}. The above details comprise what we can consider the main idea of the unified model of an AGN. This model accounts for observational differences among AGNs, which are due to the different orientations of the objects as seen from Earth and the different accretion rates and masses of the central black holes \cite{urry1995unified}.  Observations show that a blazar is an AGN with a jet of matter moving at relativistic velocities oriented near our line of sight. Blazars are the most violent AGN, emitting predominantly non-thermal radiation with strong variability across the electromagnetic spectrum \cite{blandford2019relativistic}, from the radio band to extremely high gamma-ray energies on time scales that can be as short as minutes. Among blazars, we can distinguish BL Lacertae (BL Lac) objects, which have weak or continuous featureless emission lines in the optical spectrum, and flat-spectrum radio quasars (FSRQ), which have prominent emission lines in the optical spectrum. 

The variability of blazars can be observed in different energy bands. To investigate the physical mechanisms that generate the observed variations in blazar light curves, many studies have been carried out at various wavelengths. Some techniques work from the frequency domain and others from the time domain. The main tool of analysis to use in different bands is to determine their Power Spectral Densities (PSD) \cite{vaughan2003characterizing}, since the PSD can provide clues about the mechanism driving the variability. For instance, \citet{max2014time} have modeled light curves as red noise processes with the PSD to model the variability and to set constraints on the statistical significance of interband correlations. Also, light curves and PSDs have been investigated with several other methods. Gaussian Process (GP) is especially useful for analyzing astronomical time-series data, there are even studies that have initiated more active methodological discussion on multiband time series data by implementing multi-output GP \cite{hu2020modeling}. In \citet{tarnopolski2020comprehensive} the toolset includes Lomb-Scargle Periodogram (LSP) \cite{vanderplas2018understanding}, wavelet scalogram \cite{kirby2013power}, Autoregressive Moving Average process (ARMA) \cite{moreno2019stochastic}, Continuous-time ARMA (CARMA) \cite{moreno2019stochastic}, the Hurst exponent ($H$) \cite{zywucka2020optical} and others. In fact, an algebraic relationship between the $H$-exponent of the time series and the exponent of the power-law degree distribution of the visibility graph (non-horizontal) has been matched. It has been shown that the exponent of the power-law degree distribution depends linearly on $H$ \cite{lacasa2009visibility}. $H$ measures the statistical auto-similarity of a time series, i.e. the long-range dependence or memory of a process. Small scale studies have been made to classify light curves with the $H$, where it has been found that two FSRQs and four BL Lac exhibit long-term memory in the underlying process governing the optical variability of 44 identified blazar candidates \cite{zywucka2020optical}. The overall challenge is to apply effective techniques to model the complex nature of light curve variations that occur in different bands and time scales. 

We have started a study of the variability properties in the radio band of blazars observed with a large-scale, fast cadence 15 GHz radio monitoring program with the Owens Valley Radio Observatory (OVRO) 40 m Telescope that has produced 12 years of data for over 1800 sources observed twice a week \cite{richards2011blazars}. Thus, this set of time series corresponds to the most comprehensive study of blazar variability in the radio band available at this time and is ideal for conducting our study, presenting a new perspective on the methods commonly used to study AGNs. We focus on characterizing the high-energy emission mechanisms of blazars by analyzing the variability in the radio band of the light curves of more than a thousand sources. We seek to describe the light curves, i.e., to analyze the observed flux density as a function of time of these sources as a first approximation of the complexity parameters in active galactic nuclei. We are interested in assigning degrees of stochasticity to blazars, modeling the time series of light curves as complex networks. For this purpose, we rely on visibility algorithms that convert time series into graphs, where the structure of the series is preserved in the graph topology \cite{lacasa2008time}.

HVG \cite{lacasa2012time} is a novel and direct method that can represent time series as a network according to a geometric criterion that considers the magnitude of the data and its horizontal visibility with others in the time domain. Here we use directed and undirected HVG to get degree distributions in both cases, which allows us to calculate a characteristic exponent as the degree of stochasticity, and also the Kullback-Leibler Divergence (KLD), respectively. The last one is known under a variety of names, including the Kullback–Leibler distance, cross-entropy, information divergence, and information for discrimination \cite{cover2006elements}. Meanwhile, stochastic processes also play a fundamental role in many scientific fields where we can find a dynamic in a collection of random variables evolving over time \cite{91197204}. \citet{luque2009horizontal} has demonstrated that the method we apply here efficiently discriminates randomness, and not only uncorrelated randomness from chaos, but also more complicated stochastic processes in time series can be identified, such as fractional Brownian motion. KLD is sensitive to non-evident characteristics of time series \cite{acosta2021applying}. The HVG is a method that allows us to analyze time series and its irreversibility through the calculation of the Kullback-Leibler divergence (KLD). The advantage of KLD is that, unlike other measures used to estimate irreversibility over time, KLD is statistically significant, as demonstrated by the Chernoff-Stein lemma \cite{lacasa2012time}. Moreover, in the case of astronomical observations, it is common for the data measured by telescopes to be contaminated by atmospheric (reversible) noise, yet irreversible signals continue to be well characterized by the HVG method and the KLD measurement \cite{lacasa2012time}.

In this study, we model time series of the light curves with the algorithm of the Horizontal Visibility Graph to measure a characteristic exponent to describe its stochasticity, and the Kullback-Leibler Divergence, to detect a different behavior of the light curves of blazars. We will analyze if the properties of degrees distributions are connected with the spectral classification of blazars, and we are interested in contrasting with a common measurement of variability in light curves, the excess variance. To the best of our knowledge, this is the first approach to the study of blazars using HVG, which manages to identify different ranges of KLD for different light sources. The paper is organized as follows. Section~\ref{sec:HVG} presents the HVG method and defines how we obtain the degree of stochasticity and the KLD value from the complex network. Section~\ref{sec:blazar} explains technical details about these radio observations and shows a sample of a light curve to illustrate the characteristics of data series used by this study. Section~\ref{sec:results} exposes our results after applying the method on all available blazar light curves, to finally discuss these results in Section~\ref{sec:discuss}.


\section{Horizontal visibility algorithm}
\label{sec:HVG}
\begin{figure}[H]
\includegraphics[width=\linewidth]{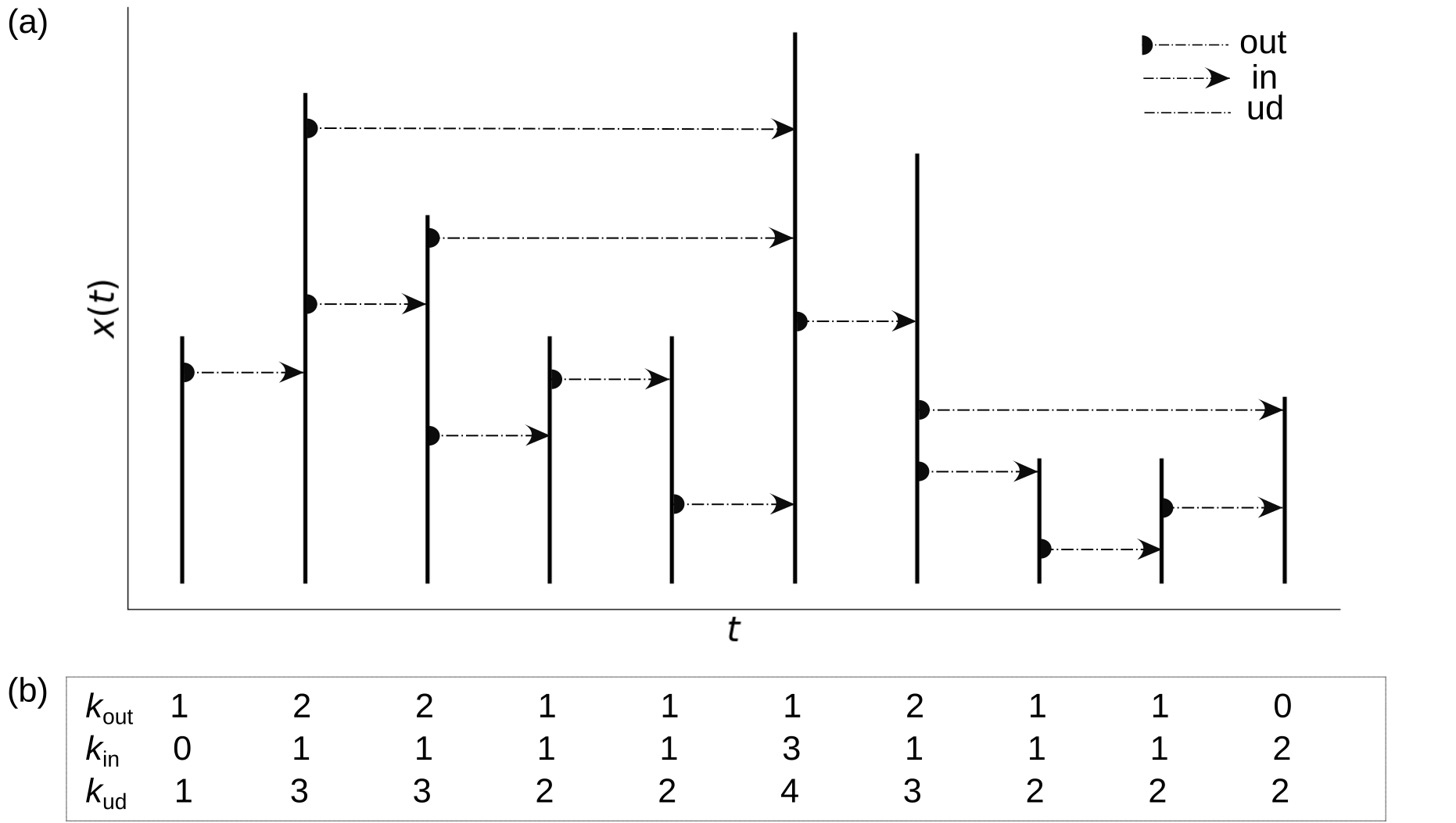} 
\caption{(\textbf{a}) A graphical description for modeling a time series with a horizontal visibility graph. Each data corresponds to a longitudinal node. (\textbf{b}) According to the visibility of each node, we can calculate the degrees $k_\text{in}$ and $k_\text{out}$ for directed HVG and $k_\text{ud}$ for undirected HVG. That is, how many times the node establishes a connection as a function of time direction and independently of this one.\label{fig:hvg}}
\end{figure}

The HVG method allows the study of dynamic systems through the characterization of their networks associated with the time series \cite{luque2011feigenbaum}. As we can see in Fig.~\ref{fig:hvg}, the HVG algorithm consists first in assigning a longitudinal node to each data in the time series. Then, depending on its magnitude, the node acquires horizontal visibility to other past and future nodes in the sense of time, this is the directed Horizontal Visibility Graph (DHVG). In that way, the number of times that the node establish an \textbf{in} and \textbf{out} connection is calculated to have degrees $k_\text{in}$ and $k_\text{out}$ for each node. With the already constructed the DHVG, we can extract the undirected version (UHVG), just considering a total degree $k_\text{ud}= k_\text{in} + k_\text{out}$. 
More details about geometrical criteria in \citet{acosta2021applying}. Thus, by counting the frequency of occurrence of each degree we obtain degrees distribution or probability distributions in the form $P(k) = n_k/n$, where $n$ is the number of data points in the time series and $n_k$ the number of nodes having degree $k$. Namely, $P=P(k_\text{ud})$ for UHVG and $P_\text{in}=P(k_\text{in})$ with $P_\text{out}=P(k_\text{out})$ for DHVG.

\subsection{UHVG to evaluate $\gamma$-exponent} 
\label{subsec:UHVG}
With UHVG any time series maps to a network with an exponential behavior for the degree distribution of the form $P(k) = \frac{1}{3}\left( \frac{2}{3}  \right)^{k-2}$\cite{luque2009horizontal}. This can be rewritten as 
\begin{equation*}
    P(k) \sim \exp\left( -\gamma_{un} k  \right),
\end{equation*}
\noindent
with $\gamma_\text{un}=\ln(3/2)\approx 0.405$, that is a limit for the uncorrelated situation proposed by \citet{lacasa2010time} to discriminate between correlated stochastic ($\gamma > \gamma_\text{un}$), or chaotic ($\gamma < \gamma_\text{un}$) processes. Here, $\gamma$ is the characteristic exponent (the $\gamma$-exponent) of the degree distribution modeled as $P(k) \sim \exp\left( -\gamma k  \right)$. The value for $\gamma_\text{un}$ was supported by analytical developments that confirmed the results provided by numerical simulations and experimental time series, but new studies show there are certain exceptions to this rule to take in consideration. \citet{ravetti2014distinguishing, zhang2017visibility} studied in depth the methodology proposed by \citet{lacasa2010time} and found several cases in which their hypothesis is not valid.

The choice of the domain for the fitted straight line of the logarithm of the probability distribution is very delicate. \citet{ravetti2014distinguishing} found that sometimes non-exponential behaviors occur, and the heavy tail of the degree distribution makes the method dependent on additional adjustments on a case-by-case basis. However, the gamma value gives useful information about the process thanks to the strengths of the HVG method that manages to maintain the intrinsic characteristics by mapping each time series conserving its properties, as long as exponential behaviors are obtained and the fitting zone is properly chosen to determine the $\gamma$-degree. Independent of the limit, it is a useful technique for a systematic analysis of long- and short-range stochastic processes with the right criteria in the fitting domain ($k$ range).

\subsection{DHVG to estimate $D$-value}
\label{subsec:DHVG}
Degree distributions $P_\text{in}$ and $P_\text{out}$ separately classify the succession between past and future events, that is, provide information about the temporal irreversibility of the associated series. At the same time, they provide a relation with the entropy production of the physical mechanism generating the series \cite{roldan2012entropy}. A rigorous way to measure the difference between two degree distributions is through Kullback-Leibler divergence, a statistical measure of "distinguishability" \cite{lacasa2012time} to quantify the degree of temporal irreversibility. The KLD between two probability functions is defined as \cite{cover2006elements}
\begin{equation*}
    D[P_{\text{out}}(k)||P_{\text {in}}(k)] = \sum _k P_{\text{ out}}(k)\log \frac{P_{\text {out}}(k)}{P_{\text{ in}}(k)}.
\end{equation*}
It is always positive and vanishes if and only if $P_\text{in} = P_\text{out}$. As $D$ moves away from zero, the distance between the distributions increases and with it the irreversibility of the series. Since it is not symmetrical, it is not a real distance measurement. Also, there are some cases in which $D\to \infty $ when $P_\text{in} = 0$. We must take into consideration that some events are unseen, especially when dealing with observational data with short duration and gaps. While the presence of gaps is not a problem for the HVG, it is not prudent to assume every event as absolutely impossible. Therefore, we reassign a new, very low, probability when it is zero. The cases where $P_\text{out}=0$ are contained in the definition itself. So, to cover the other cases, we smooth the probabilities \textbf{in} assigning a probability less than the minimum $P_\text{in}^\text{min}$ in the form $P_\text{in}^\text{min}/n$ when for certain $k$ the probability be zero, and we subtract this new probability from the others to rescale. These considerations will allow us to compare degrees between spectral classes of blazars.


\section{Blazar light curves}
\label{sec:blazar}
The blazar subclasses, BL Lac and FSRQ, are defined by the properties of their optical spectra. The spectra of FSRQs show broad emission lines, while BL Lacs show very weak or no emission lines \cite{blandford2019relativistic, hovatta2019relativistic}. Other properties of the sources are also correlated with the subclasses as described in the references above. In this article, we focus on working with observational data of blazars in the radio band.

The OVRO data were obtained from the OVRO 40 m Telescope Monitoring Program \cite{richards2011blazars}. The telescope uses off-axis dual-beam optics in which the beamwidth (FWHM) is 157 arc seconds. The cryogenic receiver uses a HEMT amplifier and is centered at 15 GHz with 2 GHz equivalent noise bandwidth. Gain fluctuations and atmospheric and ground contributions are removed with the double switching technique where one of the beams is always pointed at the source. Details of the observation and data reduction \cite{richards2011blazars} cover the absolute calibration and the uncertainties, which include both the thermal fluctuations in the receiver and systematic errors that have been added under a rigorous procedure \cite{richards2011blazars}.

It is important to study the amplitude of variability in AGN light curves. The importance of variability lies in the fact that, being a unique property of blazars, it can be used as a tool to distinguish them from other astrophysical objects \cite{zywucka2020optical}. One of the quantities most commonly used to estimate this property is the excess variance. According to \citet{turner1999x}, the normalized excess variance is
\begin{equation*}
    \sigma_\text{rms}^2 = \dfrac{1}{n\mu^2} \sum _{i=1}^n \left[ \left( X_i - \mu  \right)^2 - \sigma _i^2 \right],
\end{equation*}
designating the flux density for the $n$ points in each light curve as $X_i$, with its arithmetic mean $\mu$ and errors $\sigma_i$. The excess variance is useful for comparing the variability in different light curves. As a first application of complex networks to radio light curves, we use $\sigma_\text{rms}$ to contrast with measures that also distinguish the amplitude of the light curve variability, now according to its visibility (explained in Section~\ref{sec:HVG}). Note that we use the square root of the excess variance, which is also known as the fractional root mean square variability amplitude \cite{edelson1990broad, rodriguez1997steps}. Both quantities give us the same information, but the last one is a linear statistic that gives it in percentage terms \cite{vaughan2003characterizing}.

\begin{figure}[H]
\includegraphics[width=\linewidth]{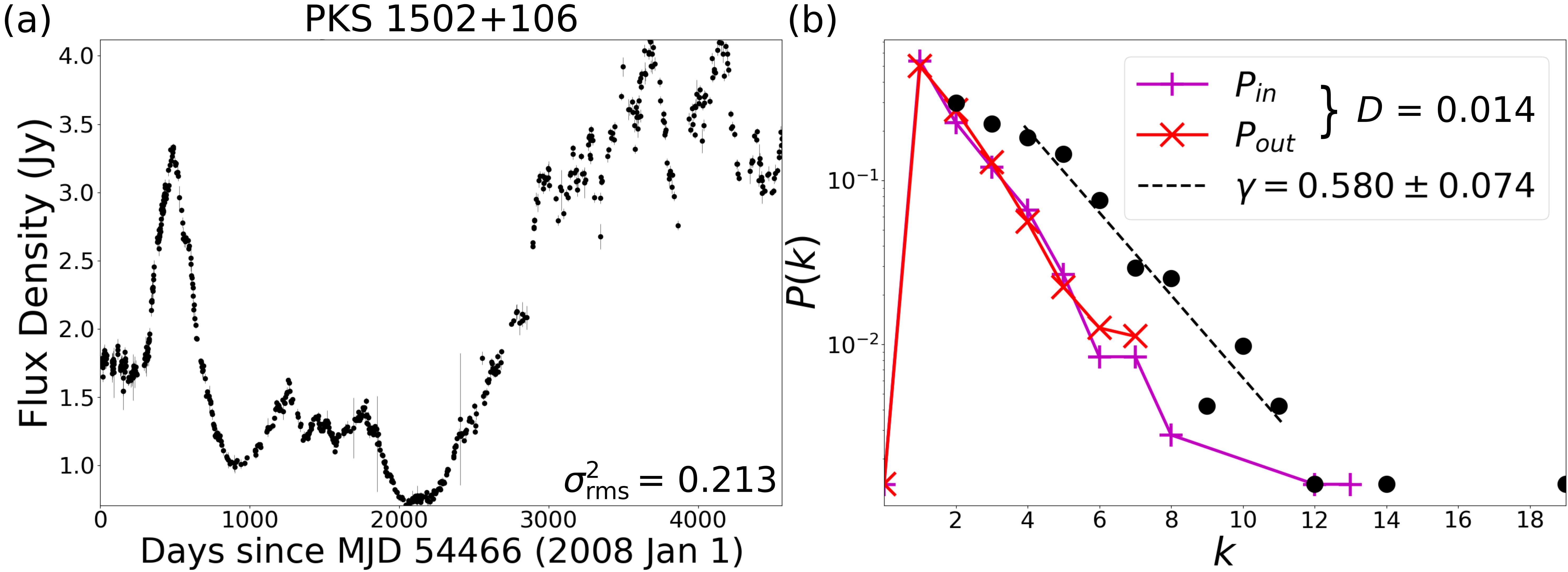} 
\caption{(\textbf{a}) Example radio light curve for the blazar PKS 1502+106 \cite{hovatta2021association}, a FSRQ source. (\textbf{b}) Semilog plot of degree distributions from DHVG, $P_\text{in}$ in magenta and $P_\text{out}$ in red, and from UHVG, $P$ in black. The value of excess variance $\sigma^2_\text{rms}$ of the light curve, $D$, from the distance between degree distributions \textbf{in} and \textbf{out}, and $\gamma$, from the exponential behavior $P(k)\sim \text{e}^{-\gamma k}$, are shown. \label{fig:light_curve}}
\end{figure}  
\unskip

\section{Results}
\label{sec:results}

We analyze the physical properties that could be conditioning the behavior of the light curves using DHVG and UHVG. Initially, we work with a data set that contains 1298 sources, where 400 are BL Lac and 898 are FSRQ. The optical classes are taken from the Roma BZCAT Multi-Frequency Catalog of Blazars 5th edition. The Multi-Frequency Catalogue of Blazars is one of the most complete lists of Active Galactic Nuclei whose emission properties are recognized as typical of blazars. It includes the list of sources and an essential compilation of multifrequency data from radio to gamma rays \cite{massaro2014multifrequency}.

We applied the algorithm of HVG over the time series of the light curves of blazars finding an exponential behavior on the degree probability distributions in most light curves, as is shown in Figure~\ref{fig:light_curve}b) for both methods UHVG (black dots) and DHVG (color solid lines). Once we calculate the degree probability distribution $P(k)$ of the UHVG, we proceed to compute the critical exponent $\gamma$ as the slope of the semilog plot with the method of least squares in a range from $k=3$ to some $k$ with the lowest probability, without including it and avoiding the low probability floor in the heavy tails (see Figure~\ref{fig:light_curve}(b) in Section~\ref{subsec:UHVG}). When making this adjustment we find that some degree probability distributions did not fit an exponential, so, we discarded those light curves. Just 5\% of BL Lac and 6.1\% of FSRQ are discarded because they deviate from the fit (some selected and discarded cases in Figures~\ref{fig:app_FSRQ_s}, \ref{fig:app_BLLac_s},  \ref{fig:app_FSRQ_d} and \ref{fig:app_BLLac_d}).

\begin{figure}[H]
\includegraphics[width=\linewidth]{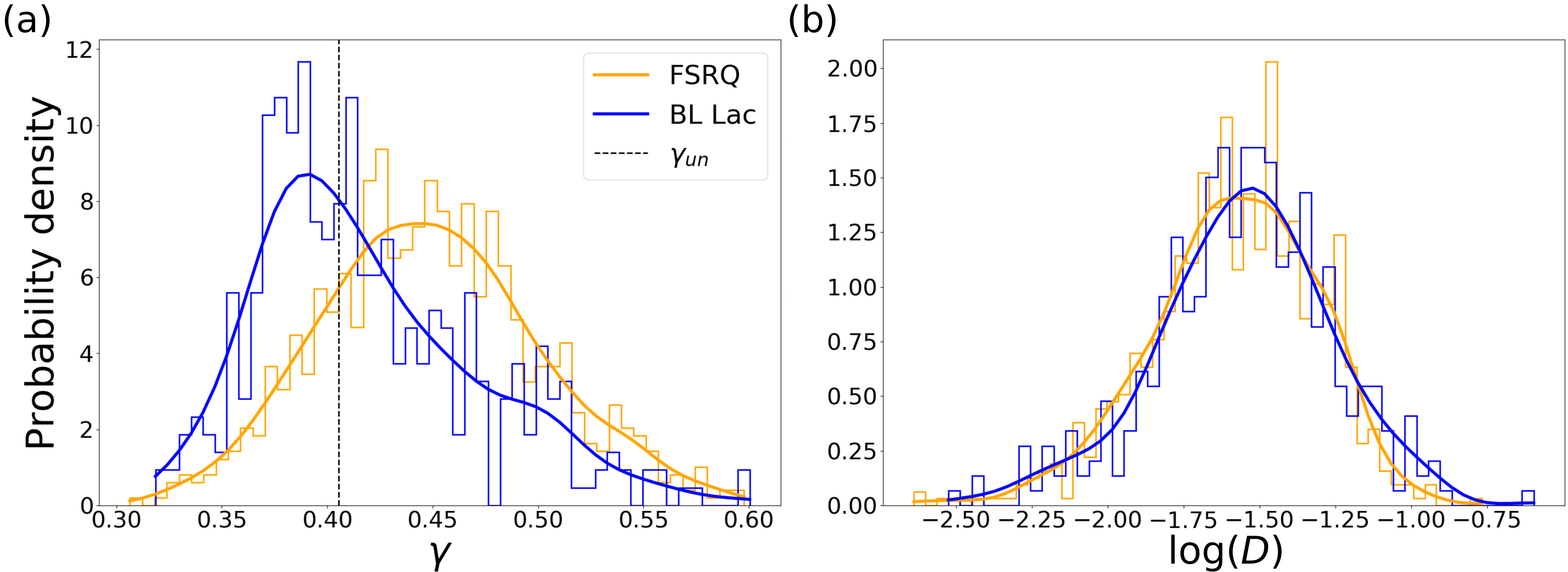} 
\caption{Probability density function (PDF) of (\textbf{a}) $\gamma$ and (\textbf{b}) $\log(D)$ values for the different subclasses of blazars. Yellow for FSRQ and blue for BL Lac. There are 843 FSRQ and 380 BL Lac sources. With dotted line in (\textbf{a}) we mark the limit $\gamma_\text{un}\approx 0.405$ between correlated stochastic and chaotic time series \cite{lacasa2010time}. \label{fig:histogram}}
\end{figure}

Now, having the well-adjusted light curves, we obtain the critical exponent $\gamma$ (UHVG) per source and we plot the PDF for these data sets as is shown in Figure~\ref{fig:histogram}a). The statistical detail of those PDFs analyses is in Table~\ref{tab1}. Figure~\ref{fig:histogram}a) shows a clear difference between the light curves of BL Lac and the FSRQ. In the dotted line, we mark the limit proposed by~\citet{lacasa2010time} between stochastic time series and chaos for the UHVG analysis. From Figure~\ref{fig:histogram}a) we observe when plotting the PDFs, the curves separate around this limit. That is, the BL Lac sources have a peak of the $\gamma$ PDF on the left of the dotted line, with $\gamma = 0.392$, whereas the FSRQ sources show the peak of the $\gamma$ PDF on the right of the limit $\gamma_\text{un} \approx 0.405$, with $\gamma = 0.446$. Thus, most BL Lac light curves show a chaotic behavior while most FSRQ light curves show a time series with a correlated stochastic behavior. 

As a second analysis, we calculate the KLD using the technique explained in Section 2.2, thus the results obtained for $D$ correspond to a smoother weighting of the original values, now it is possible to analyze cases of high irreversibility without necessarily assigning an infinity. Figure~\ref{fig:histogram}b) shows the PDF for the value of $D$, which is a measure of the irreversibility of the time series. In this case, both classes of light curves show the same behavior of this parameter, low values of $D$ between $10^{-3}$ and $10^{-1}$ as is shown if Table~\ref{tab2}, but the peaks of these PDF do not show a difference between BL Lac and FSRQ.

\begin{figure}[H]
\includegraphics[width=\linewidth]{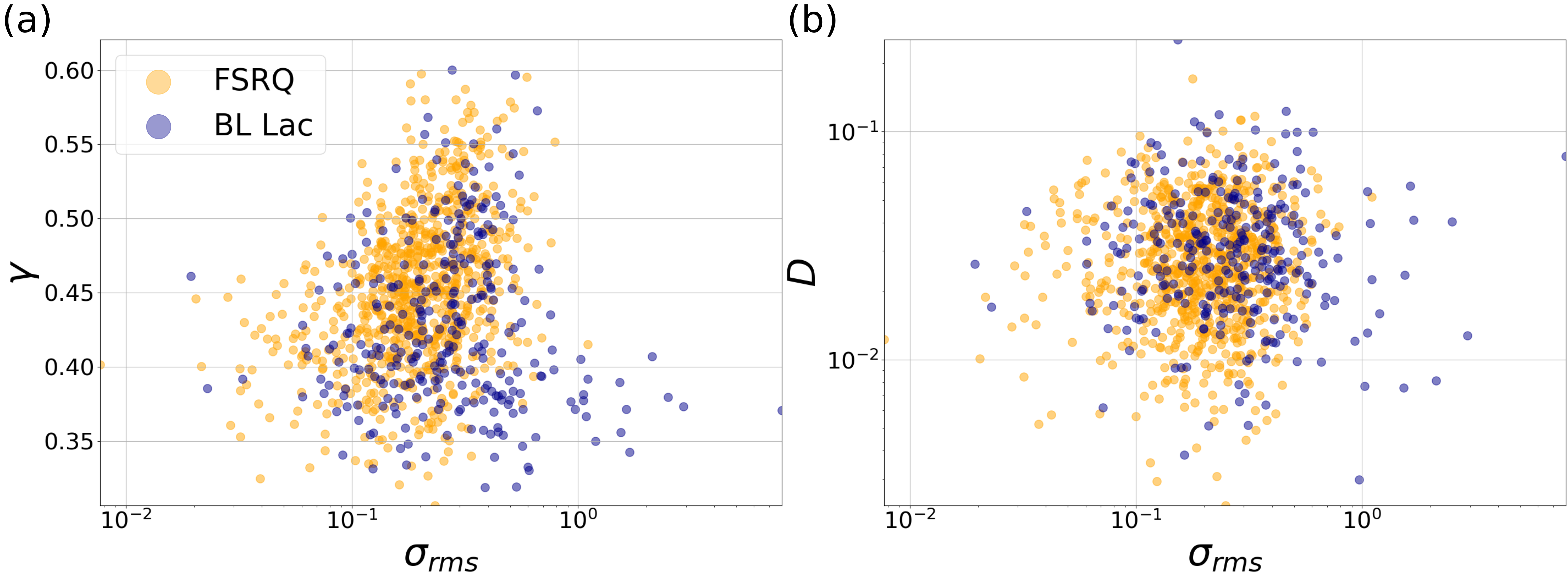}
\caption{Scatter plots of square root of the excess variance vs. (\textbf{a}) $\gamma$ and (\textbf{b}) $D$. Yellow for FSRQ and blue for BL Lac. There are 843 FSRQ and 380 BL Lac sources.  \label{fig:scatter}}
\end{figure}

Also, in order to find possible correlations between parameters of the complex network extracted from these time series and a physical measure of variability in AGN, we compare the values of $\sigma_\text{rms}$, that is the root square of the excess variance~\cite{turner1999x}, and the critical exponent $\gamma$ in Figure~\ref{fig:scatter}a) and with the $D$ value in Figure~\ref{fig:scatter}b), all dimensionless quantities. 

\unskip

\begin{table}[H] 
\caption{Statistical information on the $\gamma$ values from Figure~\ref{fig:histogram}a), i.e., the peaks of PDFs, the mean, median, standard deviation, minimum and maximum values, and the 25\%, 50\% and 75\% percentiles of the sample. \label{tab1}}
\newcolumntype{C}{>{\centering\arraybackslash}X}
\begin{tabularx}{\textwidth}{CCCCCCCCCC}
\toprule
& & & & $\gamma$ & exponent & & & & \\
\toprule
\textbf{class}  & \textbf{peak}	& \textbf{mean}	& \textbf{median}		& \textbf{std} & \textbf{min} & \textbf{25\%} & \textbf{50\%} & \textbf{75\%} & \textbf{max}\\
\midrule
FSRQ		& 0.446		& 0.449     & 0.447			& 0.052			& 0.306			& 0.413			& 0.447			& 0.482			& 0.598\\
BL Lac      & 0.392		& 0.419     & 0.409			& 0.053			& 0.319			& 0.381			& 0.409			& 0.451			& 0.600\\
\bottomrule
\end{tabularx}
\end{table}
\unskip

\begin{table}[H] 
\caption{Statistical information on the $D$ values from Figure~\ref{fig:histogram}b), i.e., the peaks of PDFs, the mean, median, standard deviation, minimum and maximum values, and the 25\%, 50\% and 75\% percentiles of the sample.\label{tab2}}
\newcolumntype{C}{>{\centering\arraybackslash}X}
\begin{tabularx}{\textwidth}{CCCCCCCCCC}
\toprule
& & & & $D$ & value & & & & \\
\toprule
\textbf{class}  & \textbf{peak}	& \textbf{mean}	& \textbf{median} & \textbf{std} & \textbf{min} & \textbf{25\%} & \textbf{50\%} & \textbf{75\%} & \textbf{max}\\
\midrule
FSRQ        & 0.026		& 0.031		&0.027  	& 0.019			& 0.002			& 0.018			& 0.027			& 0.041			& 0.170\\
BL Lac      & 0.030		& 0.034		& 0.029			& 0.025			& 0.003			& 0.019			& 0.029			& 0.043			& 0.253\\
\bottomrule
\end{tabularx}
\end{table}
\unskip

\unskip

\section{Discussion}
\label{sec:discuss}
We have applied the method of DHVG and UHVG to a sample of 1298 light curves measured from blazars. From this analysis, we find an exponential behavior on the degree probability distribution $P(k)$ for most studied sources Figure~\ref{fig:light_curve}b). We compute the critical exponent $\gamma$ from the $P(k)$ in the UHVG noting that the degree distributions are capable of detecting differences in the spectral classification of blazars as is shown in Figure~\ref{fig:histogram}a). In fact, the division between the peak reached by the PDF of $\gamma$ for the BL Lac is on the left of the $\gamma_\text{un}$ limit meanwhile the peak reached by the PDF of $\gamma$ for the FSRQ sources is on the right of this limit. That difference suggests a chaotic behavior in the time series for the BL Lac sources and a correlated stochastic behavior in the time series for the FSRQ sources. This result indicates that the distribution of the degrees, i.e., how the flux density is distributed in time is not the same for the light curves sources studied. So, the distribution of the degree $k$ of the classes of light curves is not the same and it seems that the critical exponent $\gamma$ from the exponential adjustment could be useful to distinguish between these two types of blazars. On the other hand, when measuring the irreversibility of the time series with the DHVG, the distance $D$ does not have the necessary sensibility to identify the two types of light sources, Figure~\ref{fig:histogram}b).

In Figures~\ref{fig:scatter}a) and~\ref{fig:scatter}b) we plot the critical exponent $\gamma$ (UHVG) versus the square root of the excess variance and the same for parameter $D$ (DHVG), in order to find a correlation between complex networks and a variability parameter of blazars. The excess variance is a quantity that indicates the observed relative strength of the variability of an astronomical source. Figure~\ref{fig:scatter}a) shows a slow tendency for blazar classes, whereas the KLD and the excess variance do not seem to exhibit a significant correlation in Figure~\ref{fig:scatter}b). However, as a first approach to the study of blazars with HVG, in Figure~\ref{fig:scatter}b) we managed to identify different ranges of $D$ for different sources (between $10^{-3}$ and $10^{-1}$). On the other hand, it is recommended that for better comparison between sources, the excess variance should be calculated using observations of the same duration \cite{turner1999x}. This data set contains time series from 259 to 1140 data points. Therefore, even if we could truncate the light curves to match the shortest observation, this would considerably reduce our data set and analysis. However, these quantities may not be the best parameters to consider. 

Other ways to analyze the complex networks and their degrees distribution $P(k)$ also can be considered. For instance, in many cases it is useful to consider also the complementary cumulative distribution function or CDF of a variable with a power-law distribution \cite{clauset2009info,zhang2017visibility}. However, here we have systematically obtained exponential distributions represented by the $\gamma$ exponent. Thus, here we have focused on the use of the HVG as a method to distinguish the different light curves, and to tackle this purpose we have used the HVG and analyzed the distribution of the values of the $\gamma$. In addition, it is known that when using the HVG, any random series results in a network with a degree distribution of exponential type, and it has been suggested that this is a universal feature \cite{munoz2021analysis}. If exponential forms are not obtained, the series are related to non-randomness \cite{luque2009horizontal}. Nevertheless, it would be interesting to explore with CDF to obtain perhaps a more robust statistical ﬁtting as discussed in the study of \citet{zhang2017visibility}. We will continue the analysis using other tools from complex networks analysis, as well as other physical parameters such as the PSD of the flux density time series. Although we did not find a clear correlation between $\gamma$ or $D$ and the excess variance, the PDF of the critical exponent of the degree probability distribution does show a clear difference between the two blazar classes. Thus, our results may open a new framework for the study of blazars, in which complex networks may be a valuable alternative tool to study AGNs according to the variability of their flux density.

\vspace{6pt} 

\authorcontributions{Conceptualization, B.A., W.M., D.P and P.S.M.; methodology, B.A., W.M., D.P and P.S.M.; formal analysis, B.A., W.M., D.P and P.S.M.; investigation,  B.A., W.M., D.P and P.S.M.; writing---original draft preparation, B.A., W.M., D.P and P.S.M.; writing---review and editing,  B.A., W.M., D.P and P.S.M.; All authors have read and agreed to the published version of the manuscript.}

\funding{This research was funded by ANID Chile, through FONDECYT grant number 1191351.}

\dataavailability{Restrictions apply to the availability of these data. Data was obtained from OVRO program and are available \url{https://sites.astro.caltech.edu/ovroblazars/} with the permission of Anthony Readhead, principal investigator.} 

\acknowledgments{This research has made use of data from the OVRO 40-m monitoring program \cite{richards2011blazars}, supported by private funding from the California Institute of Technology and the Max Planck Institute for Radio Astronomy, and by NASA grants NNX08AW31G, NNX11A043G, and NNX14AQ89G and NSF grants AST-0808050 and AST- 1109911. W.M. gratefully acknowledges support by the ANID BASAL projects ACE210002 and FB210003, and FONDECYT 11190853. We would like to thank Hocine Cherifi and the organizers of the ``Complex Networks 2021'' Hybrid Conference, for a great and fruitful scientific meeting, in which an early version of this work was presented.}

\conflictsofinterest{The authors declare no conflict of interest.}

\abbreviations{Abbreviations}{
The following abbreviations are used in this manuscript:\\

\noindent 
\begin{tabular}{@{}ll}
HVG & Horizontal Visibility Graph\\
UHVG & Undirected Horizontal Visibility Graph\\
DHVG & Directed Horizontal Visibility Graph\\
KLD & Kullback-Leibler Divergence\\
AGN & Active Galactic Nuclei\\
OVRO & Owens Valley Radio Observatory\\
BL Lac & BL Lacertae\\
FSRQ & Flat-Spectrum Radio Quasars

\end{tabular}}

\appendixtitles{no} 
\appendixstart
\appendix
\section[\appendixname~\thesection]{}
\begin{figure}[H]
\includegraphics[width=\linewidth]{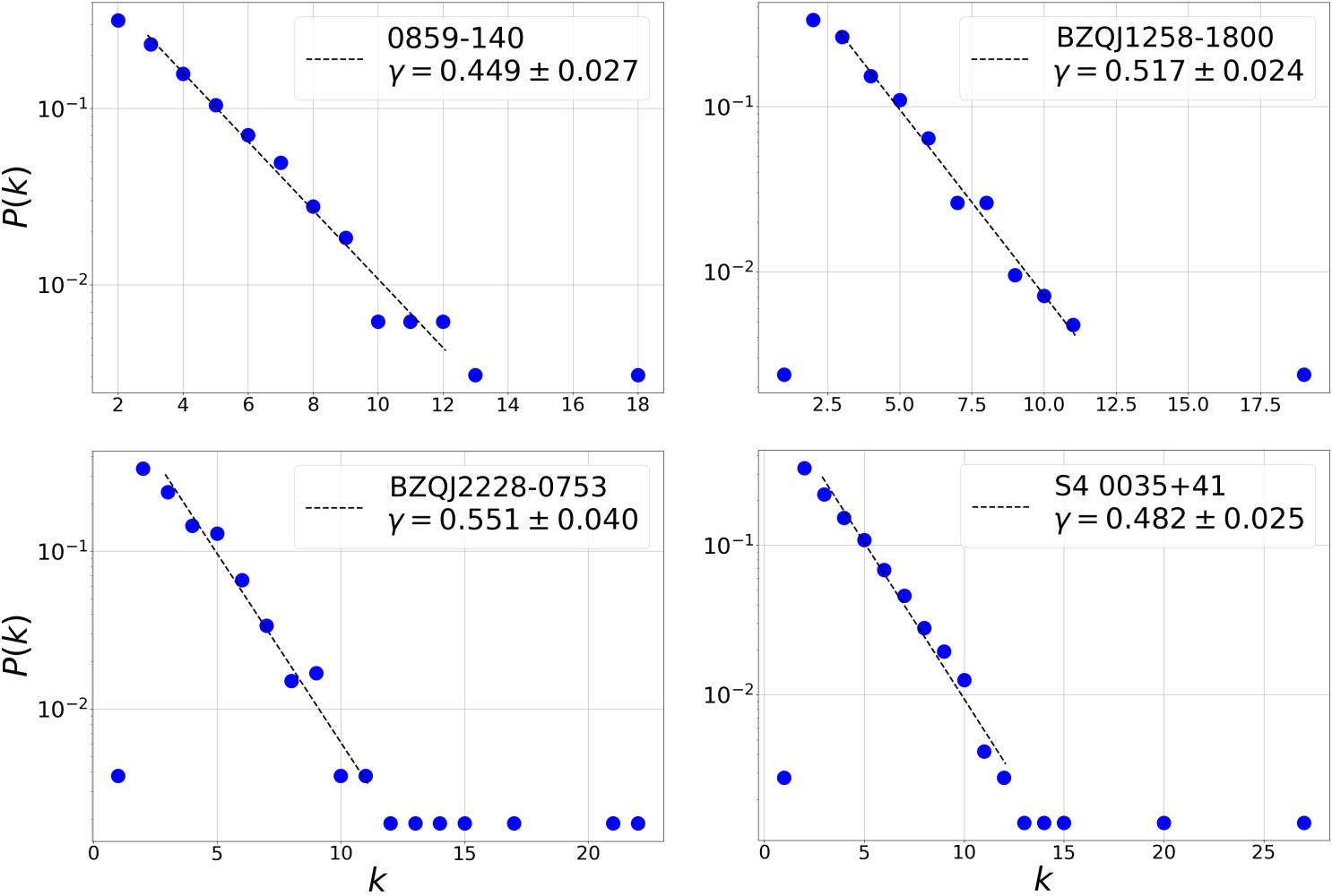} 
\caption{Examples of selected probability distribution fits to FSRQ light curves. \label{fig:app_FSRQ_s}}
\end{figure}

\begin{figure}[H]
\includegraphics[width=\linewidth]{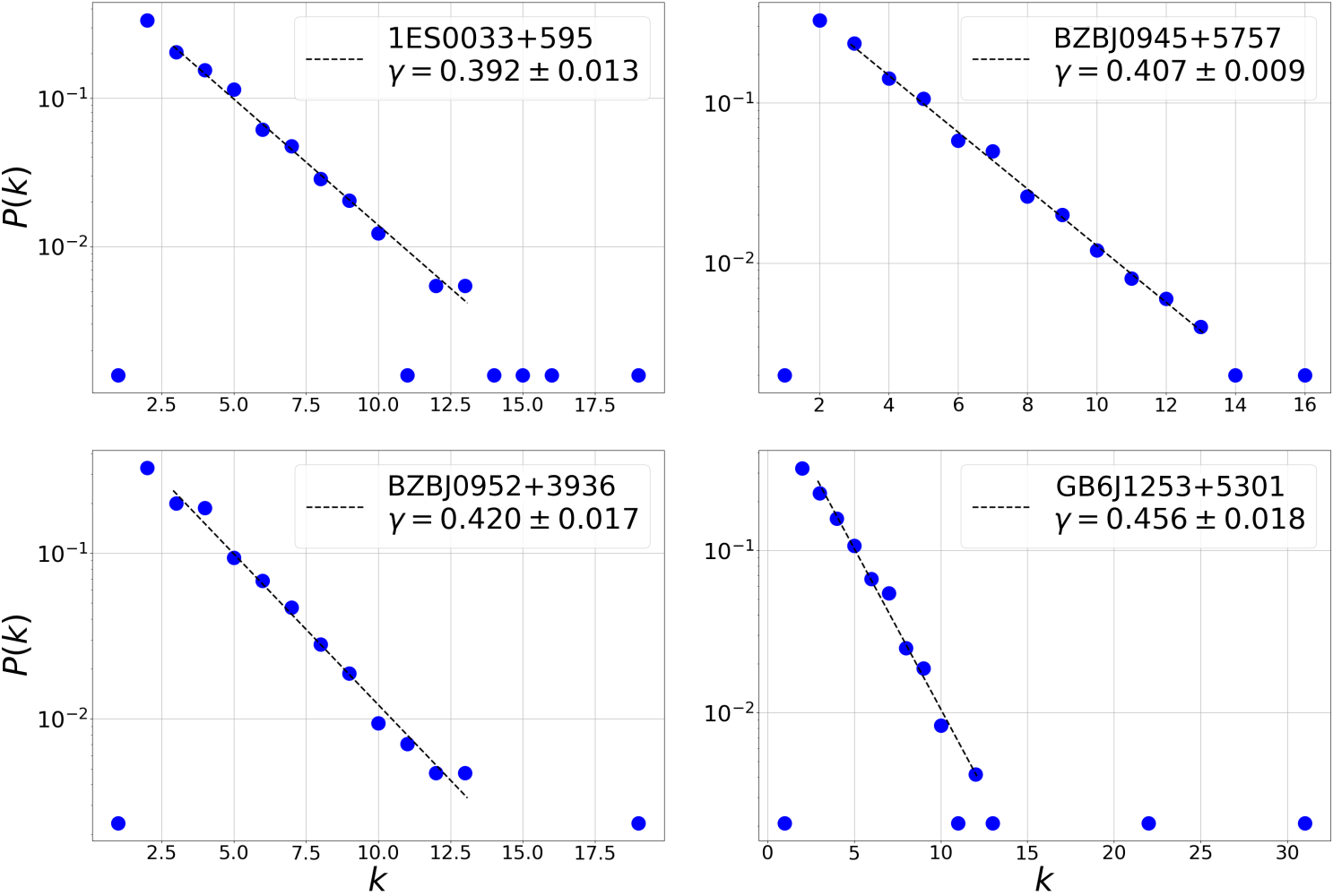} 
\caption{Examples of selected probability distribution fits to BL Lac light curves. \label{fig:app_BLLac_s}}
\end{figure}

\begin{figure}[H]
\includegraphics[width=\linewidth]{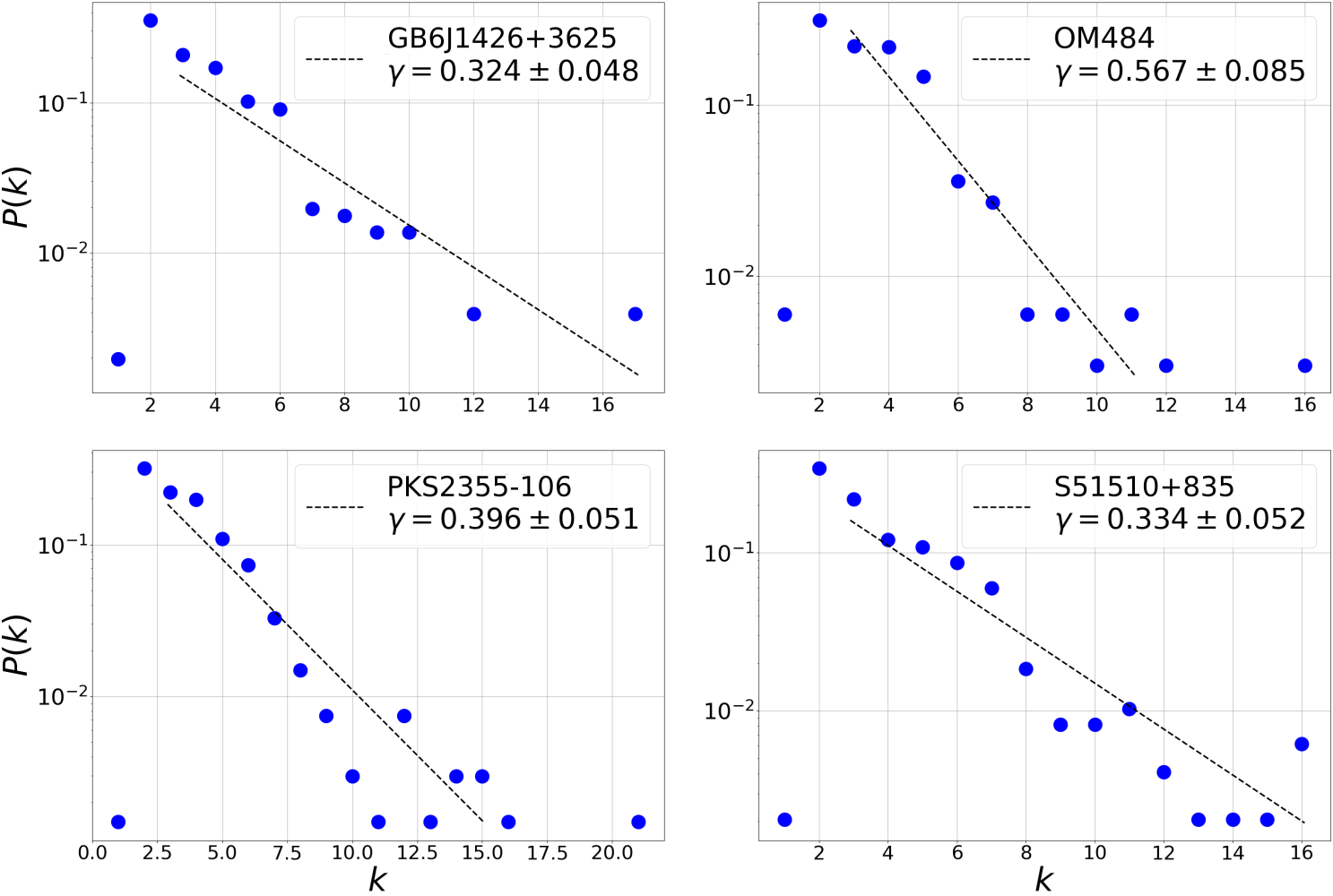} 
\caption{Examples of discarded probability distribution fits for FSRQ light curves. \label{fig:app_FSRQ_d}}
\end{figure}

\begin{figure}[H]
\includegraphics[width=\linewidth]{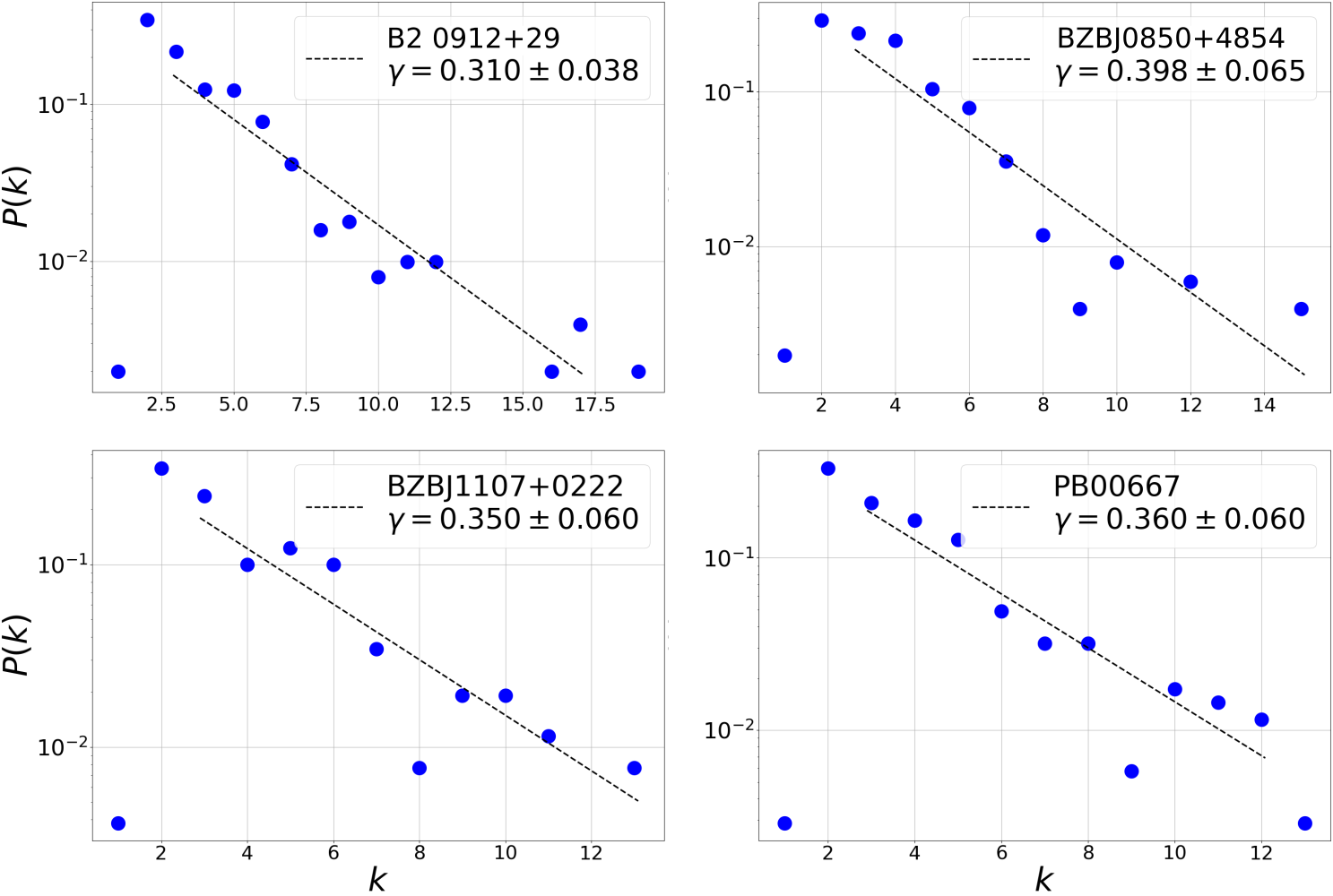} 
\caption{Examples of discarded probability distribution fits for BL Lac light curves. \label{fig:app_BLLac_d}}
\end{figure}

By crossing the OVRO source list and the Roma BZCAT Multi-Frequency Catalog of Blazars, we start with a data set that contains 1298 sources, where 400 are BL Lac and 898 are FSRQ. As from the fit (explained in Section~\ref{sec:results}) for the slope in the semilog plots for $P(k)$ according to obtain the $\gamma$-exponent from the exponential form $P(k)\sim \text{e}^{-\gamma k}$ (described in Section~\ref{subsec:UHVG}), we proceeded to select source by source such that its degree distribution was represented in this exponential behavior. Figures~\ref{fig:app_FSRQ_s} and \ref{fig:app_BLLac_s} show four of the 843 selected FSRQ sources, and four of the 380 selected BL Lac sources, respectively. Figures~\ref{fig:app_FSRQ_d} and ~\ref{fig:app_BLLac_d} show four of the 55 discarded FSRQ sources, and four of the 20 discarded BL Lac sources, respectively, in which some characteristics escape from the adequate behavior to assign a representative gamma value to the curve. Thus, for the results shown in the Figures~\ref{fig:histogram} and \ref{fig:scatter}, we used 843 FSRQ and 380 BL Lac sources.

\unskip

\unskip

\unskip

\unskip

\begin{adjustwidth}{-\extralength}{0cm}

\reftitle{References}

\end{adjustwidth}
\end{document}